\documentclass[pra,twocolumn,superscriptaddress]{revtex4-1}
\usepackage{graphics,epsfig ,color,graphicx} 

\begin{document}
\date{\today}
\title{Dynamics at Threshold in Mesoscale Lasers} 
\author{T. Wang}
\affiliation{ Universit\'e de la C\^ote d'Azur, Institut de Physique de Nice,
1361 Route des Lucioles, 06560 Valbonne, France}

\author{H. Vergnet}
\affiliation{\'Ecole Normale Sup\'erieure de Lyon, 46 All\'ee d'Italie, 69364
Lyon cedex 07, France}

\author{G.P. Puccioni}
\affiliation{Istituto dei Sistemi Complessi, CNR, Via Madonna del Piano 10,
I-50019 Sesto Fiorentino,   Italy}

\author{G.L. Lippi}
\email{Gian-Luca.Lippi@inphyni.cnrs.fr}
\affiliation{ Universit\'e de la C\^ote d'Azur, Institut de Physique de Nice,
1361 Route des Lucioles, 06560 Valbonne, France}


\begin{abstract}
The threshold properties of very small lasers (down to the nanoscale) are a topic of active research in light of continuous progress in nanofabrication.  
With the help of a simple rate equation model we analyze the intrinsic, macroscopic dynamics of threshold crossing for (Class-B) lasers whose response is adequately well--described by interplay of the intracavity photon number and the population inversion (energy reservoir).  We use the deterministic aspects of the basic rate equations to extract some fundamental time constants from an approximate analysis of laser dynamics in the threshold region.  Approximate solutions for the population inversion and for the field intensity, up to the point where the latter reaches macroscopic levels, are found and discussed.  The resulting timescales characterize the laser's ability to respond to perturbations (external modulation or intrinsic fluctuations in the lasing transition region).  Numerical verifications test the accuracy of these solutions and confirm their validity.  The predictions are used to interpret experimental results obtained in mesoscale lasers (VCSELs) and to speculate about their extension to nanolasers.
\end{abstract}
\pacs{INSERT PACS}

\maketitle

\section{Introduction}

Recent results obtained in mesoscale lasers~\cite{Wang2015} have shown the existence of unexpected dynamics in the threshold region.  Measurements conducted on a micro-VCSEL (Vertical Cavity Surface Emitting Laser) have highlighted the presence of a regime where the lasing onset is not gradual, contrary to what one would expect from the predictions of traditional models, but rather the result of unsustained self-switching, which gradually lead into continuous wave (cw) operation~\cite{Wang2015}.  This breaks away with the usual picture of laser threshold, based on a static representation, where the field becomes coherent as soon as the pump rate exceeds its threshold value -- a picture which rests on the validity of the assumption of an infinitely large system (i.e., the thermodynamical limit~\cite{Degiorgio1970,Dohm1972} for the laser), which is in essence very well satisfied by a large class of existing devices:  in practice all solid state lasers (even microdisks), gas lasers, high power lasers, etc.~\cite{note6}.

The dynamics in this intermediate region, which bridges the gap between the entirely incoherent emission and the pure lasing action, have traditionally received little attention, as this regime was difficult to reliably access in large-sized lasers, due to its narrowness and to the ensuing dominant role of technical noise (temperature and pump fluctuations, external perturbations).
Furtermore, in Class A~\cite{Tredicce1985} lasers -- whose behavior is modeled by the evolution of the sole photon number -- the dynamics of threshold crossing does not reserve particular surprises due to the restricted (one-dimensional) phase space:   in such systems the evolution of the field intensity takes a monotonic form and the only interesting aspects cover the delay time associated with the threshold crossing due to the time-dependence of the control parameter~\cite{Mandel1997,Scharpf1987}.

The larger phase space associated with the physics of the Class B~\cite{Tredicce1985} laser -- which requires the coupled evolution of photon number and population inversion for a correct description of its behavior --, instead, renders the dynamics nontrivial, since it allows for a non-monotonic evolution of the laser intensity, in addition to the appearance of an intrinsic time delay, superposed to the one induced by the bifurcation~\cite{Mandel1997,Tredicce2004}.  This intrinsic delay stems from the fact that: (1) the field intensity cannot grow until the population inversion has reached its threshold value, and (2) a sufficient amount of inversion is needed to allow for the growth of the photon number.  This introduces a causal element which requires the population to grow by a sufficient amount for the field intensity to approach its above-threshold value; it is also the cause for the appearance of a timescale proportional to the geometric mean of the two relaxation constants ($\gamma$, for the population inversion, and $K$, for the field intensity).  In the absence of this causal component, one should have expected the field intensity to grow at its own rate $K$, once the population inversion has reached its threshold value.

Nonetheless, these remarks would have remained purely threoretical, had it not been for the advent of very small devices, possessing an extended transition region~\cite{Bjork1991,Yokoyama1992} and good stability -- thanks to robust engineering of commercially available, high-quality devices~\cite{Ulm} -- which now make it possible to ask questions on the dynamical properties of the transition region.  In turn, the information gained in microdevices offers the possibility for extrapolation to nanolasers, for which there is an ever growing interest~\cite{Strauf2006,Ulrich2007,Wiersig2009,Hostein2010,Khajavikhan2012,Wu2015,Hayenga2016,Pan2016}, coupled to the need for complementary information, due to the inherent measurement bottleneck (extremely low photon flux) and to modeling difficulties stemming from the strong coupling to the environment of a limited but nonnegligible number of emitters (e.g., quantum dots).  All these systems are, by necessity (strong miniaturization and large gain), semiconductor-based and fall therefore in the Class B, the focus of this paper.

Besides the fundamental interest in a deeper understanding of the device's behaviour during lasing onset, there are important application--related aspects which may benefit from a good description of these dynamics.  The first and foremost is data transmission where the information is encoded in the optical beam by direct modulation of the laser pump~\cite{Sattar2015,Wang2015b}.
Direct modulation has been extensively studied in all kinds of semiconductor-based lasers, including VCSELs, and has become a standard textbook topic~\cite{Agrawal1993}.  However, the requirement for good quality data encoding has driven the investigations towards regimes well away from threshold -- the parameter region studied in this paper.  Modulating micro- and nanolasers in the intermediate threshold region, instead, may hold some promise~\cite{Wang2015b}, since, among others, it would allow one to exploit the low-bias regime with its accompanying low-power dissipation. 

This paper focusses on the incipient coherent dynamics arising from threshold crossing in a Class B laser~\cite{Tredicce1985} with an eye at obtaining estimates for the characteristic timescales on which the dynamics evolves and, by extrapolation, on the spectral features to be expected from the spontaneously-driven dynamics, to be compared to experimental results~\cite{Wang2016a}.  

While the complete dynamical description of the self-spiking regime which bridges the gap between the pump parameter range where the photon output is dominated by spontaneous emission, and the one where the emitted field is fully coherent requires a stochastic model~\cite{Puccioni2015}, the general properties are still described by the deterministic phase space (cf. e.g., Fig. 7 in~\cite{Wang2015} for a reconstruction, both numerical and experimental, of the phase space).  Thus, we can extract useful information on the incipient dynamics at threshold crossing from the dynamical characteristics as described by the traditional deterministic laser models.

From the minimal model for a single mode, Class B laser~\cite{Siegman1986} we compute the time needed by the population inversion to reach threshold, from an arbitrary initial condition, then using an expansion and setting a reference level for the field intensity, we obtain an estimate of the time needed to attain a macroscopic field as a function of the laser characteristic constants (relaxation rates for population inversion and field intensity).  This estimate is compared to the expression for the Relaxation Oscillations (ROs) and to direct numerical integrations of the rate equations.  This way, we gain information on the minimal time needed for obtaining the coherent field and, by extrapolation, we find the highest frequency at which the laser can respond to modulation:  this frequency will broadly correspond to the spectral component of the spontaneous excitation when the laser is biased near threshold.  Extrapolating from micro- to nanolasers~\cite{Wang2016b}, we can therefore infer some information on the behaviour of nanolasers in their broad transition region between entirely incoherent and fully coherent output. 

Section~\ref{VCSELpars} will detail the general calculations of Section~\ref{predictions} (whose limits of validity are tested in Section~\ref{num}) and Section~\ref{exptres} analyses experimental results obtained in the threshold region in small lasers either with fixed bias (as in~\cite{Wang2015}) or in the presence of pump modulation.

\section{Dynamical threshold crossing}\label{predictions}

As already proven by the comparison between experimental and numerical predictions carried out in a micro-VCSEL~\cite{Wang2015}, it is possible to obtain the essential dynamical information from a minimal laser model without resorting to the full ensemble of details which describe semiconductor-based devices.  Thus, we base our calculations on standard rate equations, written here in normalized form~\cite{Lippi2000}:
\begin{eqnarray}
\label{rateeqI}
\frac{d I}{ d t} & = & - K (1 - D) I \, , \\
\label{rateeqD}
\frac{d D}{d t} & = & - \gamma [(1+I) D - P] \, ,
\end{eqnarray}
where $I$ stands for the electromagnetic (e.m.) field intensity, $D$ for the population inversion, $K$ for the intensity losses, and $\gamma P$ represents the pump rate (i.e., power supplied to the laser).

This normalized  version of the rate equations immediately highlights the existence of the two time scales:  the e.m. field intensity evolves over a timescale $\tau_I \sim \frac{1}{K}$ (eq.~(\ref{rateeqI})), while the population inversion's timescale is $\tau_D \sim \frac{1}{\gamma}$ (eq.~(\ref{rateeqD})). 

Recalling, for convenience, the main properties of this form of rate equations, we find two sets of steady states (cf. , e.g.,~\cite{Narducci1988})
\begin{eqnarray}
\label{ss}
\left(
\begin{array}{c}
\overline{I} = 0\\
\overline{D} = P\\
\end{array}
\right) 
& \quad , \quad & 
\left(
\begin{array}{c}
\overline{I} = P-1\\
\overline{D} = 1\\
\end{array}
\right) 
\end{eqnarray}
with laser threshold value $P_{th} = 1$ (i.e., $\overline{I} = 0$).  From the linear stability analysis of the above-threshold, stable ($P \ge 1$) solution~\cite{Narducci1988,Mandel1997}  we find the eigenvalues:
\begin{eqnarray}
\label{eigenv}
\lambda & = & \frac{1}{2} \left[ - \gamma P \pm \sqrt{\gamma^2 P^2 - 4 \gamma K (P-1)} \right] \, ,
\end{eqnarray}
where the square root takes imaginary values as soon as 
\begin{eqnarray}
P & \gtrsim & 1 + \frac{1}{4} \frac{\gamma}{K} \, .
\end{eqnarray}
Given that $\gamma \ll K$ for all Class B lasers, the eigenvalues, eq.~(\ref{eigenv}), are (almost) always complex above threshold, and represent a (damped) oscillation with angular frequency 
\begin{eqnarray}
\label{omegar}
\omega_r \approx \sqrt{\gamma K (P-1)} \, ,
\end{eqnarray}
which we have obtained by neglecting the term $\gamma^2 P^2$, very small compared to $4 \gamma K (P-1)$ for all practical values of $P$.

In order to define the threshold crossing in the simplest way, we consider a pump in the form of a step function going from below to above threshold $P(t=0^-) \equiv P^- < 1$ and therefore $I(t=0) = 0$ and $D(t=0) \equiv D_0 = P^-$, according to the first set of steady-state values, eq.~(\ref{ss}).  Assuming a Heaviside function shape for the pump ($P(t<0) \equiv P^- < 1, P(t>0) \equiv P^+ > 1$), the model reduces to
\begin{eqnarray}
\frac{d I}{d t} & = & 0 \, , \\
\frac{d D}{d t} & = & -\gamma (D - P^+) \, ,
\end{eqnarray}
with the initial conditions specified above.  Thus, the e.m. field intensity remains constant, at zero, while the population inversion starts growing exponentially according to
\begin{eqnarray}
\label{initialD}
D(t) & = & (P^+ - D_0) ( 1 - e^{- \gamma t} ) + D_0 \, ,
\end{eqnarray}
which holds until the instant $\tilde{t}$ at which the population inversion reaches the other solution, eqs.~(\ref{ss}):  $D(\tilde{t}) = 1$.  Starting from this instant, the first of the model equations (\ref{rateeqI}) acquires a positive right-hand-side (r.h.s.) and the e.m. field intensity starts growing~\cite{foot1}.  

The value of $\tilde{t}$ can be straightforwardly obtained from eq.~(\ref{initialD}):
\begin{eqnarray}
D(\tilde{t}) & = & 1 = (P^+ - D_0) ( 1 - e^{- \gamma \tilde{t}} ) + D_0 \, , \\ 
\label{t-tildedef}
\tilde{t} & = & - \frac{1}{\gamma} \log \left[ \frac{P^+-1}{P^+-D_0} \right] \, ,
\end{eqnarray}
where we are assured that $\tilde{t} > 0$ by the fact that $0 < P^+-1 < P^+ - D_0$.  Starting from this instant, the full model, eqs.~(\ref{rateeqI}-\ref{rateeqD}), must be used in its nonlinear form and no closed solution exists for the time evolution of the physical variables.  However, if we concentrate on the initial phases of the e.m. field intensity growth, we can gather some information on the timescale over which the laser turns on.

Assuming that the laser intensity $I$ is small in its initial phases ($I \ll 1$), we can suppose its influence on the evolution of $D$ to be negligible (since $(1+I) \approx 1$ in the r.h.s. of eq.~(\ref{rateeqD})) and obtain an approximate form for $D(t > \tilde{t})$:
\begin{eqnarray}
D(t) & = & D(\tilde{t}) + D(\delta t) \, , \\
\label{Dabovethr}
& = & 1 + (P-1) \left( 1 - e^{- \gamma \, \delta t} \right) \, , \qquad \delta t \equiv t- \tilde{t} \\
& \approx & 1 + (P-1) \left[1 - (1 - \gamma \, \delta t) \right] \, , \\
\label{approxD}
& = & 1 + (P-1) \gamma \, \delta t \, ,
\end{eqnarray}
which holds as long as $\delta t \ll \frac{1}{\gamma}$, a condition which is very well satisfied in practice (and which can be checked {\it a posteriori} -- cf. section~\ref{num}).  In the last set of equations we have dropped the superscript in the pump value, since from here on it is unambiguous.

We can now use this approximate solution to get an approximate solution for the initial phases of the e.m. field intensity growth by replacing $D(t)$ from eq.~(\ref{approxD}) into the rate equation for $I$ (eq.~(\ref{rateeqI})), which can be recast as:
\begin{eqnarray}
\frac{d (\log I)}{d t} & = & \gamma K (P-1) \delta t \, .
\end{eqnarray}

Direct integration provides
\begin{eqnarray}
\label{formalintegral}
\int_{\tilde{t}}^t d(\log I) & = & \gamma K (P-1) \int_{\tilde{t}}^t  (t^{\prime} - \tilde{t}) d t^{\prime} \\
& = & \frac{1}{2} \gamma K (P-1) (t - \tilde{t})^2 \, ,
\end{eqnarray}
which only holds until a time $t < t_M$, to be determined.  The left-hands-side  of eq.~(\ref{formalintegral}) provides $\log I(t)$ up to a constant ($\log I(\tilde{t})$) which corresponds to a mathematical divergence, since $I(\tilde{t}) = 0$.  Besides being unphysical, this is an artefact of the model, which considers only the deterministic evolution of the coherent fraction of the e.m. field:  spontaneous emission is not included in this description.  A self-consistent solution can only be obtained by including the spontaneous photons in the description, but the complexity of the model increases considerably;  the average properties of the lasing transition, however, are still correctly given by the set of rate eqs.~(\ref{rateeqI}-\ref{rateeqD}). 
Thus, we can use the correct physical condition (i.e., the average value of the spontaneous emission in the lasing mode at threshold) to estimate the value of $I(\tilde{t})$, thus avoiding the unphysical divergence.  Indicating with $I_0$ this value (i.e., the value of $I$ at $t = \tilde{t}$), we obtain the approximate expression for the e.m. field intensity growth:
\begin{eqnarray}
\label{approxI}
I(t) & = & I_0 e^{\frac{1}{2} \gamma K (P-1) (t - \tilde{t})^2} \, ,
\end{eqnarray}
which already provides us with a wealth of (deterministic) information about the initial phases of the growth of the laser intensity:
\begin{itemize}\itemsep0cm
\item[$\bullet$] the growth is exponential but with a quadratic time dependence -- since the solution~(\ref{approxI}) holds only at short times, the quadratic growth in time indicates a slower initial growth than what would result from a linear time dependence;
\item[$\bullet$] the speed at which the laser intensity grows depends on the distance of the pump from threshold ($P-1$):  the larger the pump, the faster the growth; 
\item[$\bullet$] the time constant for the intensity growth is not proportional to $K^{-1}$, as one would, mistakenly but intuitively, expect from the timescale evolution of the intensity (cf. eq.~(\ref{rateeqI})), but rather to the geometric mean of the two time constants $(\gamma K)^{-1/2}$;  
\item[$\bullet$] the actual time constant for the exponential growth is 
\begin{eqnarray}
\label{expgrowth}
\tau_{exp} = \sqrt{\frac{2}{\gamma K (P-1)}} \, ,
\end{eqnarray}
i.e., the distance of the pump to threshold induces a hyperbolic lengthening of the time scale, with the usual divergence, typical of critical slowing down~\cite{Haken1983}, taking place as $P \rightarrow 1$.
\end{itemize}
While intuitively unexpected, the dependence of the timescale on the product of the two relaxation constants for the physical variables is logical.  Indeed, it is not sufficient for the e.m. field intensity to grow at a rate $K^{-1}$ since the population inversion must have the time to increase as well in order for the photon number to develop.
Notice that the relaxation oscillations, eq.~(\ref{omegar}), appearing around the above-threshold solution (thus, far beyond the intensity ranges we are considering here) have the same parameter dependence as the time delay $\Delta t \equiv t - \tilde{t}$, apart from a numerical coefficient.  This point is significant since it shows how the time constants appearing in all parts of the transient evolution are closely related to each other by the intrinsic physical interplay between the laser variables.

The limits of validity of the solution we have obtained for the transient can be established in the following way.  The transient dynamics of class B lasers is characterized by a delay in the laser intensity growth, accompanied by an overshoot of its value beyond its asymptotic state with damped oscillations~\cite{Agrawal1993}.  In a phase space representation, this dynamics correspond to a trajectory spiraling into the fixed point~\cite{Lippi2000}.  Strong deviations for the growth of the population inversion from the approximate solution, eq.~(\ref{approxD}), are expected, and numerically found, when the laser intensity exceeds its asymptotic value $\overline{I} = P-1$.  Thus, we can set the limit of validity at a fraction of this value $a \overline{I}$ ($a<1$, arbitrary) to determine the maximum time value $t_M$ for which the approximate solution for $I(t)$ (eq.~(\ref{approxI})) holds:
\begin{eqnarray}
a (P-1) & = & I_0 e^{\frac{1}{2} \gamma K (P-1) (t_M - \tilde{t})^2} \, ,
\end{eqnarray}
which immediately gives an estimate for $t_M$:
\begin{eqnarray}
\label{exprtM}
t_M & = & \tilde{t} + \sqrt{\frac{2}{\gamma K (P-1)} \log \left( \frac{a (P-1)}{I_0} \right) } \, .
\end{eqnarray}
Since we are trying to estimate the time necessary to attain a fraction, $a$, of the steady state value for the laser intensity, it does not make sense to consider the limit $P \rightarrow 1$, thus the potential divergences present in the expression on the r.h.s. of eq.~(\ref{exprtM}) lie outside the realm of the interesting physical parameter ranges.

\section{Numerical verifications}\label{num}

\begin{figure}[ht!]
\includegraphics[width=0.9\linewidth,clip=true]{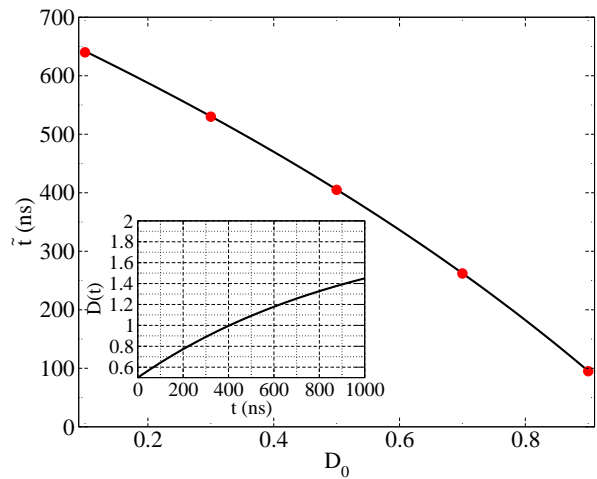}
\caption{
Comparison between the time needed by the population inversion to reach threshold ($D=1$) obtained from the direct integration of the model (eqs.~(\ref{rateeqI},\ref{rateeqD})) -- dots -- and from the approximate solution (eq.~(\ref{t-tildedef})) -- solid line --.  The inset shows the full temporal evolution of the population inversion numerically computed in the limit of zero field intensity (eq.~(\ref{initialD})) over a time interval equal to its inverse relaxation time ($\gamma^{-1}$) for an initial condition corresponding to the central point of the main figure:  $D_0 = 0.5$ -- the threshold value ($D=1$) is reached in the inset at $t = 400 ns$, matching ordinate of the central point in the main figure.  The asymptotic value, reached in infinite time, is $D = P = 2$ in this inset.  For this and the following figures the parameter values used are (unless otherwise indicated):  $\gamma = 1 \times 10^6 s^{-1}$, $K = 1 \times 10^8 s^{-1}$, $P = 2$.} 
\label{t-tilde} 
\end{figure}

A verification of the approximate solutions is obtained by comparing the analytical predictions to the numerical values resulting from the integration of the model, eqs.~(\ref{rateeqI},\ref{rateeqD}), obtained with a first-order Euler scheme programmed in GNU Octave. 
The temporal evolution of the population inversion (cf. inset of Fig.~\ref{t-tilde}) displays a growth corresponding to that of a saturating exponential, as predicted by eq.~(\ref{initialD}).  The crossing time $\tilde{t}$ ($\tilde{t}$:  $D(\tilde{t}) = \overline{D} = 1$) can be found directly from this trajectory (and more precisely from the numerical file).  We also remark that, as implicit in the previous discussion, the population inversion $D$ grows beyond its asymptotic value in the process of laser threshold crossing (cf. discussion in section~\ref{discussion}).

\begin{figure}[ht!]
\includegraphics[width=0.9\linewidth,clip=true]{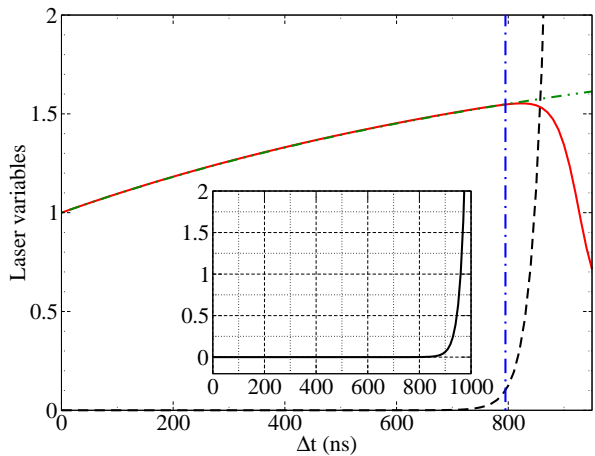}
\caption{
Temporal evolution of the laser variables as a function of time, $\Delta t = t - \tilde{t}$, starting from the instant at which the population inversion reaches threshold ($t = \tilde{t}$).  The curves represent the laser intensity (dashed black line) and the population inversion (solid red line) numerically computed from the model (eqs.~(\ref{rateeqI},\ref{rateeqD})) compared to the approximate solution (eq.~(\ref{Dabovethr})) for the population inversion above threshold (dash-double-dotted green line).  The vertical dash-dotted blue line marks the estimated maximum time $t_M$ for which the approximations hold (cf. eq.~(\ref{exprtM})):  the figure proves that the approximate (dash-double-dotted line) and the exact (solid line) solution for the population inversion give an excellent match until $t = t_M$ (and even beyond).  Inset:  Laser intensity (vertical axis) vs. time (horizontal axis) -- units as in main figure.  The (solid) line shows the shape of the field intensity ($I$) computed from the approximate solution (eq.~(\ref{approxI})):  its shape qualitatively matches quite well the true temporal evolution of the intensity (dashed line in main figure) obtained from the full model (eqs.~(\ref{rateeqI},\ref{rateeqD})) -- for the inset we have used $\gamma K/2$ in place of $\gamma K$ (cf. text for details).  All computations in this figure have been based on an initial value for the field intensity $I_0 = 1 \times 10^{-10}$ (representing the average spontaneous emission).
} 
\label{time-evolution} 
\end{figure}

Fig.~\ref{t-tilde}  shows the comparison between the time $\tilde{t}$ necessary for the population inversion to reach its above-threshold steady state value (continuous line) as predicted from eq.~(\ref{t-tildedef}),  and the numerical time obtained from the temporal trajectory (dots).  Not unexpectedly, since the approximation used in the derivation of eq.~(\ref{t-tildedef}) is very well verified, the agreement is excellent.  

Once threshold is crossed, the numerical integration has to start from a good estimate of the ``initial" value of the laser field intensity.  The deterministic rate equation model does not account for the background {\it noisy} dynamics which holds the intensity value constant (in average) around the value of the spontaneous emission.  If one starts the integration of eqs.~(\ref{rateeqI}-\ref{rateeqD}) with a deterministic initial condition (e.g., $D_0 = 0.5$, as in one of the simulations run for Fig.~\ref{t-tilde}), during the whole transient where $D(t) < \overline{D} = 1$, the laser intensity decays away to ever smaller numbers.  This is an artefact of the model and should not be mistaken for a physical effect.  Continuing the simulation from unphysically low values of the laser intensity (e.g., much lower than the average spontaneous emission level) would artificially increase the latency time needed to reach macroscopic intensity values, and thus affect the maximum values reached by the population inversion, and, as a consequence, by the laser intensity at its peak (not discussed here -- the full time evolution can be seen, for instance, in Ref.~\cite{Lippi2000}).  Thus, it is crucial to consider a reasonable estimate of the average spontaneous emission.  Traditionally, the following physical considerations have been employed to conceptually define threshold:  for the stimulated emission to overcome the spontaneous emission and concentrate on the lasing mode all (or most) of the energy, the number of photons in the lasing mode has to equal the number $N$ of modes available for the spontaneous emission (i.e., the number of electromagnetic cavity modes).  In other words, while in average the number of spontaneous photons is $\langle n_{s,j} \rangle = 1$ for each mode ($j = 1 \ldots N$), the (average) number of stimulated photons in mode $p$ must be $\langle n_{st,p} \rangle = N$ for lasing action to occur~\cite{note5}.  Thus, if we consider a laser whose cavity possesses $N$ modes, its relative average spontaneous intensity {\it at threshold} will be $\frac{\langle I_{sp} \rangle}{\overline{I}} = \frac{1}{N}$.  Without entering into details, macroscopic lasers have values of $10^7 < N < 10^{12}$ (and even beyond); small semiconductor lasers are characterised by $N \approx 10^5$, while smaller cavities exit the realm of small-sized lasers to approach the nanoscale.  A more detailed discussion, supported by stochastic calculations, can be found in~\cite{Wang2015,Puccioni2015}.  Here we use values $10^5 \le N \le 10^{10}$, specified in the figures as appropriate.

The evolution of the population inversion for a time $t > \tilde{t}$ is displayed in Fig.~\ref{time-evolution}.  The continuous lines (red online) shows the population inversion numerically integrated from the model, eqs.~(\ref{rateeqI}-\ref{rateeqD}), while the dashed line (green online) represents the predictions of the approximate expression, eq.~(\ref{initialD}).  The graph convincingly shows that the analytical approximation holds well beyond $\tilde{t}$, even once the laser intensity $I$ starts growing away from $0$.  Indeed, the two curves are superposed for times exceeding $t = 8 \times 10^{-7} s$ (for the parameter values of the figure), and remain very close until $I \approx \frac{\overline{I}}{2}$ ($\frac{\overline{I}}{2}$ = 0.5 for the chosen parameters).

\begin{figure}[ht!]
\includegraphics[width=0.9\linewidth,clip=true]{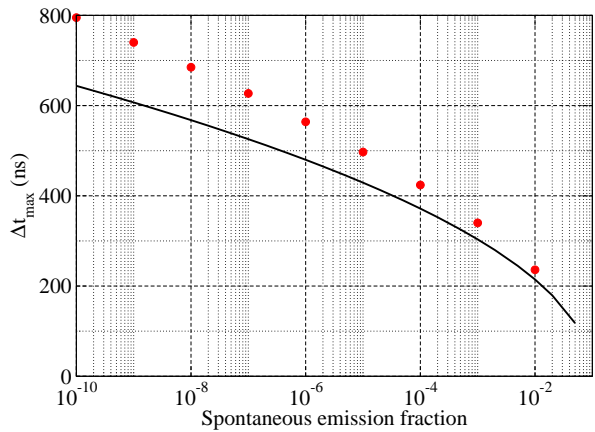}
\caption{
Examination of the degree of accuracy in the estimate of the time needed by the laser intensity to reach a reference value (defined as a fraction, $a$, of its asymptotic value, $\overline{I}$), starting from different initial values $I_0$ of the field intensity.  The latter quantity mimicks the average fraction of spontaneous emission (later defined as $\beta$-parameter and plotted here on the horizontal axis) coupled into the lasing mode in devices with different cavity volumes.  Time ($\Delta t_{max} = t_M - \tilde{t}$, plotted on the vertical axis) is measured starting from the instant $\tilde{t}$ at which the population inversion reaches threshold.  The approximate solution, eq.~(\ref{exprtM}), is plotted as a solid line, while individual time delay values obtained from the full model (eqs.~(\ref{rateeqI},\ref{rateeqD})) are plotted as dots.  The figure clearly shows that the agreement between the approximate and full solution improves as the cavity size shrinks (larger values for the spontaneous emission fraction), thus indicating that our estimates improve for ever smaller lasers.  In this figure, we are extending the analysis to $\beta = 10^{-2}$, where the error is of the order of 10\%, for the sake of the experimental discussion of mesoscale lasers (Section~\ref{expt}) and in view of an extension to nanolasers (Section~\ref{discussion}).  For this figure the reference value for the intensity has been set at $a = 0.1$, i.e. 10\% of the asymptotic value $\overline{I}$, according to eq.~(\ref{ss}) -- right hand group of solutions. 
} 
\label{deltatmax} 
\end{figure}

The analytical predictions of section~\ref{predictions} have provided also an estimate of the maximum time value for which the approximate analysis holds.  Fig.~\ref{deltatmax} shows a comparison between the estimated time, as a function of the spontaneous emission fraction (i.e., $\langle I_{sp,p} \rangle / \overline{I}$).  The agreement here is somewhat less good than the one previously obtained, due to the fact that we have retained only the linear term (first-order correction) in the expressions for the population inversion, eq.~(\ref{approxD}), to obtain an approximate behaviour for the initial phases of the intensity growth, as reproduced by eq.~(\ref{approxI}).  It is from this latter equation that we have estimated the maximum time, eq.~(\ref{exprtM}), represented as a time difference $\Delta t_{max} \equiv (t_M -\tilde{t})$ in Fig.~\ref{deltatmax}.  Notice, however, that the order of magnitude is correctly obtained and that the largest error is of the order of 20\%:  it occurs, unsurprisingly, for the lower values of the spontaneous emission, which lead to longer values of $t_M$.

We also remark that the threshold set for determining the value of $t_M$ ($a = 0.1$) falls well within the range of validity for the approximate expression of the population inversion given by eq.~(\ref{approxD}):  the time $t_M$ is marked in Fig.~\ref{time-evolution} by the vertical dot-dashed line (blue online) -- at this instant, the numerical and the analytical expression for $D(t_M)$ coincide.

Finally, we look at the shape of the initial growth of the field intensity, as displayed in the inset of Fig.~\ref{time-evolution} for comparison with the dashed line (black online) in the same figure.  The overall shape is quite well reproduced, even surprisingly well for an approximate solution with a growth rate as large as that of a quadratic exponential, but for the value of the time constant $\tau_{exp}$ which, for the sake of graphical comparison, has been doubled.  Given the rather crude approximations used to obtain the shape of the growing intensity, eq.~(\ref{approxI}), then used for estimating the time $t_M$, the qualitative agreement is quite satisfactory.  As a last remark, the value of the delay time $\tau_{exp}$ used in the numerical comparison, larger than the one coming from the analytical estimate, brings its value a bit closer to the actual response time (Fig.~\ref{deltatmax}) and to the relaxation oscillation period, estimated from the linear stability analysis, eq.~(\ref{omegar}).

\section{Interpretation of experimental results}\label{expt}

In order to apply our results, obtained for generic relaxation parameters, to small-scale (semiconductor-based) lasers, we quantitatively analyze the predictions and discuss their limits in the following subsection.  Throughout this section we will make use of the $\beta$-parameter, defined as the (average) fraction of spontaneous emission coupled into the lasing mode~\cite{Rice1994}, and will estimate the below-threshold steady-state values for $I$ and $D$ on the basis of the slightly more complex rate equation model including the average spontaneous emission (Cf. e.g.,~\cite{vanDruten2000}).

\begin{figure*}[ht!]
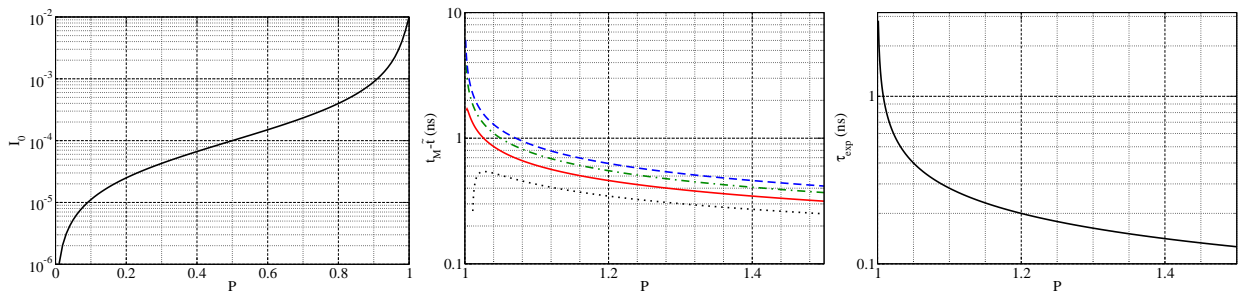

\includegraphics[width=0.3\linewidth,clip=true]{ss-SS-b-4-K1.eps}
\includegraphics[width=0.3\linewidth,clip=true]{dt-b-4.eps}
\includegraphics[width=0.3\linewidth,clip=true]{tau-exp.eps}
\caption{
Application of the approximated estimates of Section~\ref{predictions} to a $\beta = 10^{-4}$ laser.  
Left panel:  estimated initial intensity value, corresponding to the fraction of spontaneous emission, based on the expressions of~\cite{vanDruten2000} (cf. text for details) for different values of the normalized pump $P$.  Center panel: estimated maximum time for which the approximate expressions derived in Section~\ref{predictions} hold, for different values of pump (horizontal axis) and different initial values of the laser intensity -- curve legend:  dotted black line $I_0 = 10^{-3}$; solid red line $I_0 = 10^{-4}$; dot-dashed green line $I_0 = 10^{-5}$; dashed blue line $I_0 = 10^{-6}$.  The different values of $I_0$ are chosen to examine different ranges of below-threshold photon numbers (e.g., to examine the response to a periodic modulation).  Right panel:  estimated exponential time constant for the growth of the field intensity at different levels of target pump $P$ (eq.~(\ref{expgrowth})).  Parameter values used for the simulations:  $\gamma = 2.5 \times 10^9 s^{-1}$, $K = 1 \times 10^{11} s^{-1}$, $\beta = 10^{-4}$, $a = 0.1$.
} 
\label{belowthreshold} 
\end{figure*}

\subsection{Theoretical predictions for a VCSEL}\label{VCSELpars}

Close comparison with experimental results requires a reasonable quantitative estimate of the initial values of the field intensity $I_0$ (I.e., $I(t = \tilde{t})$).  Such an estimate can be obtained from a model~\cite{vanDruten2000} which accounts for the evolution of the photon number $n$ and includes the fraction of spontaneous emission $\beta$ coupled into the lasing mode.  A comparison between the predictions of that model and our current discussion can be made with the help of the following identifications: $I_0 = \beta n_0$ ($n_0$ in~\cite{vanDruten2000} representing the photon number below threshold when the pump change is applied) and where our threshold value $P=1$ corresponds to $M=1$ in~\cite{vanDruten2000}.   

The limits of validity (eq.~\ref{exprtM}) of the expansions we have used are illustrated in the center panel of Fig.~\ref{belowthreshold} for different initial values of the field intensity -- these values are sampled from the ensemble shown in the left panel of the same figure, keeping in mind the fact that we are interested in explaining experimental results obtained on a laser with estimated $\beta \approx 10^{-4}$.  We see that, unless the laser is quite close to threshold, the estimated growth exponential time constant (eq.~\ref{expgrowth}) is well within the limits of the approximation and we can use its estimate, quantitatively reproduced in the right panel of Fig.~\ref{belowthreshold}.  Keeping in mind that the exponential growth (exponent $m$) corresponds to the multiplicative factor $\mathcal{F}$ for the intensity given in Table~\ref{egrowth},
\begin{table}
\caption{\label{egrowth} Multiplicative factor for the field intensity $I$ due to the exponential growth (exponent $m$) computed on the basis of eq.~(\ref{expgrowth}).}
\begin{tabular}{c c r}
$m$ & \ \  &$\mathcal{F}$ \\
1 & & 2.7 \\
2 & & 7.4 \\
3 & & 20 \\
4 & & 55 \\
5 & & 150\\
\end{tabular}
\end{table}
we can directly determine the time needed for the intensity to get to the desired level (within the approximations of the analysis).

\begin{figure}[ht!]
\includegraphics[width=0.8\linewidth,clip=true]{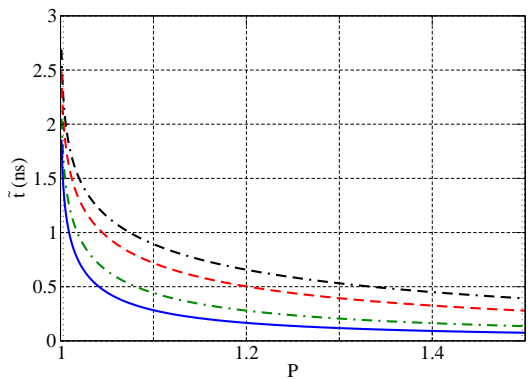}
\caption{
Time ($\tilde{t}$) required by the population inversion $D$ to grow to its threshold value ($\overline{D} = 1$) for different values of its initial condition ($D_0$) as a function of the target pump rate values $P$ (eq.~(\ref{t-tildedef})).    The different curves correspond to: double-dash-dot black line, $D_0 = 0.17$; short dashed red line, $D_0 = 0.5$; long dashed green line, $D_0 = 0.8$; solid blue line, $D_0 = 0.9$.  $\gamma_{\parallel} = 2.5 \times 10^9 s^{-1}$.
} 
\label{graph-tbar} 
\end{figure}

To the buildup time for the field intensity, we need to add the lag time needed by the population inversion to reach its steady state (i.e., threshold) value ($\overline{D} = 1$, eq.~(\ref{ss}) -- r.h.s. group of solutions) which can be obtained as follows.  First, we need to estimate the initial value of the population inversion $D_0$ to be input in eq.~(\ref{t-tildedef}); to this purpose, we use again the model which includes the spontaneous emission contribution~\cite{vanDruten2000} with its expression for the steady state population inversion:
\begin{eqnarray}
\label{ssDvD}
\overline{D} & = &  \beta \frac{\gamma_{\parallel}}{K} N_0 \, , \\
\label{ssDours}
& = & \frac{P}{1 + \overline{I}} \quad {\rm our \, notation} \, ,
\end{eqnarray}
where the r.h.s. of eq.~(\ref{ssDvD}) uses the steady state population inversion ($N_0$) of~\cite{vanDruten2000}.  After conversion into our notation (eq.~(\ref{ssDours})), this amounts to a linear relation in $P$ below threshold (i.e., $\overline{I} \ll 1$).  This result coincides with the one obtained from the simpler model we have used (Section~\ref{predictions}) and confirms its practical usefulness.  With this information, we can estimate the lag time introduced by the growth of $D$ towards threshold using eq.~(\ref{t-tildedef}), which results in the plots of Fig.~\ref{graph-tbar} for different values of $D_0$.  Thus, using the delay times plotted in Fig.~\ref{graph-tbar} and~\ref{belowthreshold} (right panel), we can estimate the total response time for the laser we are studying.

\subsection{Experimental results}\label{exptres}

The objective of this section is not to perform measurements to check the predictions, but rather to use the understanding gained from the previous analysis to interpret two phenomena which appear otherwise puzzling and which are all the more important as investigations in the transient dynamics of very small (thus large $\beta$ -- defined as the fraction of spontaneous emission coupled into the lasing mode~\cite{Rice1994}) lasers.  The first is the appearance of a {\it resonance} in the device's response subject to a modulation in the excitation, the second is the spontaneous growth of a (at first broadband) frequency component in the radiofrequency (RF) spectrum of a laser biased in the threshold region, but in the absence of relaxation oscillations of a deterministic nature.  The measurements have been performed on two nominally identical semiconductor lasers (in practice somewhat different, as is common in most devices of this kind).

\begin{figure}[ht!]
\includegraphics[width=0.9\linewidth,clip=true]{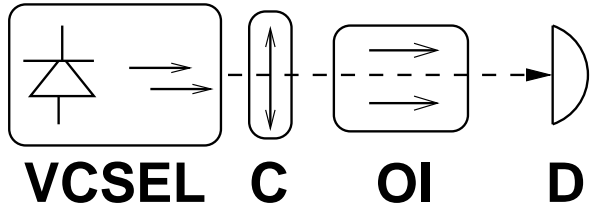}
\caption{
Schematics of principle for the optical part of the experimental setup, valid for the two experiments.  C stands for collimating optics, OI for the optical isolator and D for a photodetector.  Cf. text for details.
} 
\label{setup} 
\end{figure}

First, we consider the observations obtained from a VCSEL periodically driven across its threshold through a sinusoidal modulation of its excitation current.  The device is a VCSEL-980 (manufacturer Thorlabs) emitting at $\lambda =980 nm$ (nominal wavelength) with a maximum output power $P_{max} \approx 1.85 mW$~\cite{ThorlabsVCSEL980}.  The optical setup is schematically shown in Fig.~\ref{setup}:  the VCSEL output is collimated and passed through an optical isolator -- OI in Fig.~\ref{setup}, realized with a Polarizing Beam Splitter followed by a Quarter-Wave-Plate) -- to avoid backreflections into the source, then it is detected by a fast photodetector (Thorlabs PDA8GS) whose output is digitized with a LeCroy Wave Master 8600A oscilloscope.  The DC part of the power supplied to the VCSEL is provided by a commercial power supply, while a modulation is added through the Bias-Tee included in the TEC module (Thorlabs TCLDM9) from the sinusoidal signal generated by an HP E4421B function generator.  Additional details on the experiment can be obtained from the Supplementary Information section of Ref.~\cite{Wang2015}.  The laser threshold, though {\it ill-defined} due to the relatively large fraction of spontaneous emission coupled into the lasing mode ($\beta \approx 10^{-4}$~\cite{Wang2015}, i.e., small VCSEL), is just below, but very close to $1 mA$ for this device.

Here, we concentrate on this aspect of the observations:  the response of the laser to an external modulation for one fixed value of the bias current and one fixed modulation amplitude.  What is varied is the frequency at which the sinusoidal modulation is added to the laser.  Notice that the value of modulation amplitude is the one sent by the generator onto a $50 \Omega$ load (the nominal load of the TEC), which is larger than the actual value of current that is actually arriving at the laser junction.  Since we compare relative measurements, the actual value of injected current is not important.  Furthermore, a certain amount of attenuation is present when the modulation frequency exceeds $0.5 GHz$ (current specification -- revised by the Manufacturer from the time the TEC module was purchased).  While this perturbs somewhat the comparison, we will realize from the discussion of the results that its influence on the interpretation of the experimental outcome is only marginal.

\begin{figure}[ht!]
\includegraphics[width=0.9\linewidth,clip=true]{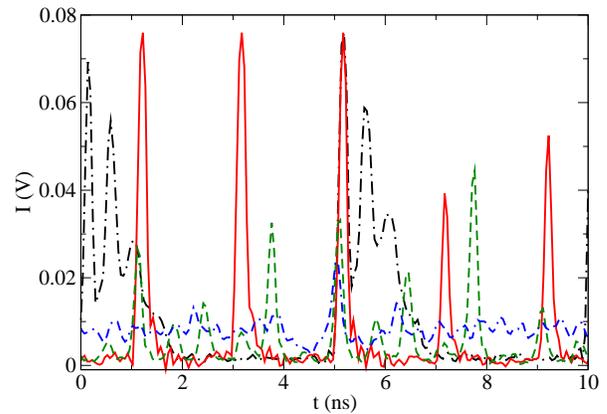}
\caption{
VCSEL response to a sinusoidal modulation at the following frequencies:  dash-dotted black line, $f = 0.2 GHz$; solid red line, $f= 0.5 GHz$; dashed green line, $f = 1.5 GHz$; double-dash-dotted blue line, $f = 2.5 GHz$.  Other parameters are given in the text.  
} 
\label{modulation} 
\end{figure}

Fig.~\ref{modulation} shows the laser response for a bias current level $i_b = 1.0 mA$ and modulation amplitude $i_m = 2 mA$ at different modulation frequencies (cf. figure caption).  Because the laser is biased very close to threshold and the modulation amplitude is large, then during the input current oscillation it is periodically driven far below threshold.  Hence, our analysis of Section~\ref{predictions} is relevant and provides an interesting basis for the interpretation of the observations.

Due to the sinusoidal shape and periodicity of the modulation, the actual values of minimum (and maximum) inversion become dependent on all parameters (in particular the frequency), but what is important is that the turn-on is characterized by a delay (not shown, as all output is triggered at the same level) and by the shape of the signal.  At the lowest frequency value shown ($0.2 GHz$) the laser has sufficient time to turn on, but, especially, to reach large enough values of the population inversion to display well-pronounced damped relaxation oscillations (dash-dotted line -- black online -- in Fig.~\ref{modulation}).  The important point is that the relaxation oscillations are not only damped, but they are also convoluted with the current modulation shape which terminates them rather ``abruptly" (cf. Fig.~\ref{modulation}).  The laser emission duty cycle (i.e., the fraction of the modulation period over which the laser emits at all) is $< 40\%$ at this frequency; since the pump is below threshold for half of the period, we can estimate the delay time to be $\sim 1 ns$.  This delay time is consistent with the estimate coming from Figs.~\ref{belowthreshold} and~\ref{graph-tbar} which would give a delay time of the order of $1 ns$ (considering five or six $\tau_{exp}$ plus $\tilde{t}$) if we assume an effective pump $P_{eff} \approx 1.5$ (cf. below for the choice of this value).

The most interesting point, however, is the frequency of the relaxation oscillation:  from the figure, we find it to be of the order of $2 GHz$.  Comparison to RF spectra for continuous and stable laser excitation (not shown here) shows that this relaxation oscillation frequency corresponds to a dc bias value $i_b \sim 1.5 mA$.  This means that, while the laser is emitting at low modulation frequency, we can consider it as operating at an {\it effective} bias at about $50\%$ above threshold. 

Increasing the frequency, we remark that at $0.5 GHz$ (solid line -- red online -- in Fig.~\ref{modulation}) the laser response consists of a single large spike with a short {\it duty cycle} ($\approx 12\%$), which is readily interpretable -- for the initial parts of the dynamics -- in terms of the delay time needed to reach threshold.  For this parameter value, the pump is above threshold for approximately $1 ns$ and since the pulse lasts approximately $0.2 ns$, then we estimate the delay to $\approx 0.8 ns$.  Here the slope of the pump is larger and therefore the growth is faster.  While we observe a partial reduction in the delay time (approximately 80\% of the previous value for a modulation frequency 2.5 times larger), the lasing pulse has become considerably shorter and now amounts to approximately half a cycle of the relaxation oscillation.  An additional increase in $f$ does not strongly affect the delay time but instead encroaches on the fraction of the period in which the laser may turn on.  As an example, we may look at the dashed line (green online) in Fig.~\ref{modulation} ($f = 1.5 GHz$) where not only the amplitude of the peaks is strongly reduced, but their repeatability is considerably affected by the fluctuations from one cycle to the next.  At the largest value of modulation frequency used ($f = 2.5 GHz$, double-dash-dotted line -- blue online -- in Fig.~\ref{modulation}) the laser does not even turn on (the attenuation introduced by the electrical cutoff plays also a role here).  

Notice that for $f = 0.5 GHz$ the intensity peak is larger than at $f = 0.2 GHz$, indicating that a larger portion of population inversion is suddenly converted into intensity (and that modulation signal coupling into the laser is efficient).  This corresponds to an apparent {\it resonance} in the response which is {\bf \em not} produced by an oscillatory nature of the system:  for a considerable portion of the modulation period the laser is below threshold and it is not susceptible to oscillations of any kind.  The {\it resonance} comes from the optimum compromise between two competing processes:  the sudden growth of the intensity due to the pump modulation and the larger encroaching of the delay time on the intensity buildup.

\begin{figure*}[ht!]
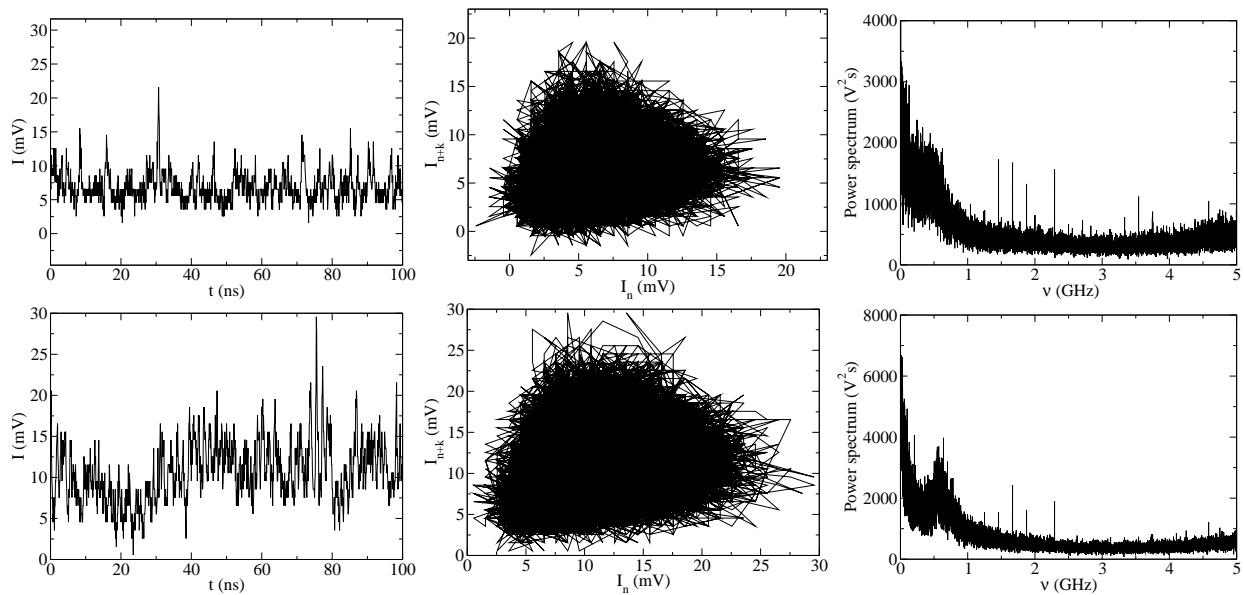

\includegraphics[width=0.3\linewidth,clip=true]{tt1-26mA2Tb.eps}
\includegraphics[width=0.3\linewidth,clip=true]{phsp1-26mA2b.eps}
\includegraphics[width=0.3\linewidth,clip=true]{psp1-26tr2B.eps}\\
\includegraphics[width=0.3\linewidth,clip=true]{tt1-30mA5Tb.eps}
\includegraphics[width=0.3\linewidth,clip=true]{phsp1-30mA5b.eps}
\includegraphics[width=0.3\linewidth,clip=true]{psp1-30tr5B.eps}\\
\caption{
Analysis of experimental signals which possess indications of a {\it repetitive} phenomenon in a mesoscale VCSEL.  
Left panels:  temporal evolution of the laser intensity measured in the VCSEL biased at $i = 1.26 mA$ (top) and $i = 1.30 mA$ (bottom).  Each figure corresponds to a trace sampled at a repetition rate $0.1 ns$ with nominal detector's bandwidth $9.5 GHz$ (cf. text for details).  The threshold for continuous operation (i.e., no pulses) is estimated for this laser at approximately $i = 1.45 mA$~\cite{Wang2015} -- although nominally equal, this physical device is not the same one on which the modulation measurements of Fig.~\ref{modulation} are obtained, thus the actual parameter values at which the different regimes appear do not coincide.  Center panels:  phase space reconstruction with embedding delay~\cite{Solari1996} equal to 10 points (i.e. $1 ns$).  No structure is recognizable indicating lack of global oscillations:  this representation confirms that the only {\it repetitive}  feature is the appearence of laser spikes, rather than continuous oscillations, which would appear as rings in the phase space representation.  Right panels:  radiofrequency power spectra of the temporal signal depicted in the left panels, computed by Fourier transform on $10^6$ points.  For the larger value of current (bottom panel) a frequency component stands out, visible, but with more difficulty, in the top panel as well.  In both cases, it corresponds to the {\it repetition frequency} of the laser spikes (poorly defined for the lower values of current -- top figures), not to regular oscillations such as the usual Relaxation Oscillations of a semiconductor laser.  The fine lines correspond to residual perturbations which are present in the absence of the optical signal.
} 
\label{pulsing} 
\end{figure*}

We now turn to the second experiment:  the analysis of the laser dynamical response for fixed bias in the threshold region.  With an experimental setup very similar to the one of Fig.~\ref{setup} (cf. Supplementary Information section in~\cite{Wang2015}) the laser, temperature--controlled and biased at fixed, very stable current values close to threshold, displays spontaneous spiking in its output (Fig.~\ref{pulsing}, left panels).  The temporal signal shows a sequence of pulses, more or less spaced depending -- in part -- on the bias current value, with no overall oscillation:  the latter is proven by the phase space reconstruction which shows a dense covering of the plane (center panels in Fig.~\ref{pulsing}) and confirms the results of the independent test using the time-delayed autocorrelation~\cite{Wang2015} which also indicates uncorrelated pulses.  A certain amount of {\it structure}, however, results from the radiofrequency power spectra (right panels in Fig.~\ref{pulsing}) which we now analyze.

The self-induced pulsing has a stochastic nature and, as such, cannot be directly explained through the approach we have adopted here (a fully stochastic technique, instead, provides an excellent match to the experimental observations~\cite{Puccioni2015}).  Nonetheless, we may use our deterministic dynamics to extract some of the salient features which result from the measurements and thereby gain a better understanding of their dynamical underpinning.  We will do so by turning around the role of fluctuations in the physical variables into those of the problem parameters (e.g., pump rate).

Let us suppose that, due to a fluctuation, the physical variables ($I$ and $D$) may at a given instant not be at equilibrium with the corresponding parameter values; to fix the ideas for a discussion, we consider a fluctuation which makes the population inversion grow.  Following our discussion of Section~\ref{predictions} we find that the laser will go through two latent periods:  the first characterized by $\tilde{t}$ (lag time for the growth of population inversion to its threshold crossing) the second by $\tau_{exp}$ (exponential growth constant for the field intensity out of noise).  On the basis of the discussion of Section~\ref{VCSELpars}, we can obtain an order of magnitude for the total waiting time $t_w = \tilde{t} + \alpha \tau_{exp}$ ($\alpha$ being a coefficient which accounts for the amount of growth out of noise).  

Since the laser is biased very close to threshold (i.e., $P \lesssim 1$), the lag time for the population inversion is negligibly small (cf. Fig.~\ref{graph-tbar}).  Thus all the delay stems from the growth of the field intensity, whose characteristic values are given in Fig.~\ref{belowthreshold} (right panel).  The steep decrease of $\tau_{exp}$ indicates that its values easily drop below $1 ns$, suggesting that two or three time constants may bring the delay time in the region of up to $2 ns$.  Such values qualitatively match the hint of a feature 
appearing in the spectrum (upper right panel) for the lower current value displayed:  around $0.5 GHz$ one recognizes the incipient growth of a peak, overshadowed towards lower frequency components by a larger spectral component.  This also matches the temporal signal, since the spikes are typically spaced farther apart than the minimal time.  

When the bias current is increased, we expect the fluctuations to become more effective and to produce with more ease (thus in larger number) intensity pulses -- their amplitude, however, will be controlled by the eigenvalues, associated to the bias point in phase space, and since variations in pump for the curves shown in the top and bottom panels are very small, one can assume the eigenvalues to be of the same order of magnitude:  thus one expects pulses of similar height.  This is in agreement with the experimental observations (left panels of Fig.~\ref{pulsing}).  The more frequent spikes at larger pump rate values (bottom row in the figure) will occur with a {\it maximal frequency} dictated by the minimal spacing originating from their duration and from the time necessary for the pulse to grow.  Once again, the time constants resulting from the analysis are compatible with a $1 GHz$ frequency component (bottom right panel in Fig.~\ref{pulsing}) since $\tau_{exp}$ decreases very rapidly for $P>1$ (in the range of $1 ns$ and below -- Fig.~\ref{belowthreshold}).

As remarked in the discussion of the experimental results obtained with pump modulation, no oscillations involving the two laser variables ($I$ and $D$) are at play here:  the broadband frequency components which are displayed in the RF spectrum are the result of a sequence of intensity spikes brought about by the transient growth of the field intensity at threshold thanks to fluctuations in the presence of gain very close to the conditions for laser self-oscillation (in the sense of the regenerative field dynamics, rather than in the coupled population--intensity oscillation characteristic of ROs).  

Finally, we can speculate on the applicability of the predictions obtained in this paper to nanoscale devices.  While for the moment direct measurements of nanolaser dynamics are technically impossible (insufficient photon flux for current detectors and digitization and storage electronics), nonetheless, we have numerical evidence for the persistence of the observed threshold dynamics throughout the nanoscale, at least for $\beta$ values up to $0.1$~\cite{Wang2015}  (Supplementary Information).  Similarities in the behaviour between meso- and nanoscale lasers are expected also in the presence of pump modulation~\cite{Wang2015b}.  Thus, the considerations on the response time scales and on the {\it pseudo-resonance} experimentally observed in our mesoscale device may also apply to nanodevices, for which the temporal scaling is favourable to a threshold modulation since the intrinsic time scales can become considerably faster.  The present findings may pave the way for reliable modulation around threshold at sufficiently high rates, thus permitting low-dissipation (thanks to very low bias) modulated laser output.

\section{Discussion and conclusions}\label{discussion}

The simple analysis presented in this paper provides us with predictions about the turn-on time of the laser when switched up from below (and close to) threshold.  Quite understandably, this time is composed of two parts:  a lag time needed by the population inversion to reach threshold (Eq.~(\ref{t-tildedef})), and a second delay time required by the buildup of the coherent field.  Both depend on the above--threshold pump level, while the lag time also depends on the initial condition.  What is particularly interesting is the identification of a short--term characteristic growth time constant (eq.~(\ref{expgrowth})) for the field intensity, whose expression bears a close resemblance to the expression for the above-threshold relaxation oscillations:  $\frac{\tau_{exp}}{T_r} \approx \frac{1}{\pi \sqrt{2}}$.  We thus find a link between the main feature which characterizes the phase space when lasing is well established and the growth constant away from the unstable (below-threshold) solution~\cite{Lippi2000,Lippi2000prl}.  In spite of the approximate nature of the quantities we derive, their range of validity is found to be relatively large and numerical tests show that they account for a sizeable part of the transient dynamics -- the remainder (which requires a fully nonlinear analysis and is best found from direct numerical simulations) takes place so fast as to play a marginal role in the estimate of resonse time constants.

These analytical findings have been applied to the interpretation of experimental results obtained on a mesoscale laser in two different regimes:  one where the pump is periodically modulated across the threshold region with large amplitudes, the second one where the laser is biased very close to threshold and displays intrinsic dynamics.  Both regimes are found to be characterized by time scales which are in agreement with those obtained from our analysis, which helps explaining the observed dynamics.  
Additional examples of the application of this analysis to the threshold region in microlasers will be shown in forthcoming papers.

In summary, in this paper we have obtained approximate expressions for the times at which the population inversion reaches its threshold and the field intensity attains macroscopic values, together with approximate solutions for both variables within the time intervals just defined.  The agreement is quite satisfactory in all cases (and even excellent in some), in spite of the extreme simplicity of the analysis.  These considerations allow for a deeper insight into the threshold crossing properties of Class B lasers at the microscale and, by extrapolation, permit inferences which hold for nanoscale devices~\cite{Wang2016b}.

\end{document}